\documentclass[aps,prl,twocolumn,showpacs,superscriptaddress,10pt]{revtex4-1}
\usepackage{graphicx} 
\usepackage{amssymb} 
\usepackage{amsmath}
\usepackage{mathrsfs}
\usepackage{soul,color}

\begin{document}

\setstcolor{red}

\title{Sequential terahertz pulse generation by photoionization and coherent transition radiation in underdense relativistic plasmas}

\author{J. D{\'e}chard}
\email{jeremy.dechard@cea.fr}
\author{A. Debayle}
\author{X. Davoine}
\author{L. Gremillet}
\author{L. Berg\'e}
\affiliation{CEA, DAM, DIF, F-91297 Arpajon, France}

\date{\today}

\begin{abstract}
Terahertz (THz) emission by two-color, ultrashort optical pulses interacting with underdense helium gases at ultrahigh intensities ($> 10^{19}\,\mathrm{W/cm}^2$) is
investigated by means of 3D particle-in-cell simulations. The THz field is shown to be produced by two mechanisms occurring sequentially, namely, photoionization-induced radiation (PIR) by the two-color pulse and coherent transition radiation (CTR) by the wakefield-accelerated electrons escaping the plasma. For plasmas of atomic densities~$> 10^{17}\,\mathrm{cm}^{-3}$, CTR proves to be the dominant process, providing THz bursts with field strength as high as $100\,\mathrm{GV/m}$ and energy in excess of $1\,\mathrm{mJ}$. Analytical models are developed for both
the PIR and CTR processes, which correctly reproduce the simulation data.
\end{abstract}

\maketitle

In recent years, the generation of terahertz (THz) radiation by ultrashort laser pulses has stirred much interest due to many applications in medecine and security
\cite{Tonouchi:np:1:97}. Among other techniques \cite{Chan:rpp:70:1325,Vicario:prl:112:213901,Stepanov:ol:33:2497}, frequency conversion through a plasma
spot seems particularly promising given the absence of emitter damage and the production of intense broadband fields \cite{Cook:ol:25:1210,Kim:np:2:605}.
In laser-gas interactions at moderate pump intensities ($\sim 10^{14-15}\,\mathrm{W/cm}^2$), various mechanisms come into play, depending on the intensity level and
the temporal laser profile. While THz radiation by single-color laser pulses appears mainly mediated by the longitudinal ponderomotive force through transition-Cherenkov emission \cite{Damico:njp:10:013015}, transverse photocurrents prevail when using temporally asymmetric two-color pulses
\cite{Kim:np:2:605,Babushkin:njp:13:123029, Andreeva:prl:116:063902}. This trend has been verified up to sub-relativistic intensities $\lesssim10^{18}\,\mathrm{W/cm}^2$
\cite{Debayle:pra:91:041801,Gonzalez:sr:6:26743}.

Moderate pump intensities routinely supply less than $10\,\mu\mathrm{J}$ THz yields \cite{Oh:apl:102:201113}, so that progress remains to be done for producing $\mathrm{mJ}$-level
pulses with $\sim \mathrm{GV/m}$ field strength, which could be helpful for remote sensing applications. Ultrahigh intensity (UHI) lasers appear well suited in this regard
because of their ability to generate strong charged particle currents. In thin solid foils irradiated at intensities $>10^{19}\,\mathrm{W/cm^2}$, high-energy ($\sim 500\,\mu\mathrm{J}$)
THz pulses associated with high conversion efficiencies ($\eta \sim 5\times 10^{-4}$) have been reported and ascribed to either transient electron/ion currents at the target rear
surface \cite{Gopal:prl:111:074802} or coherent transition radiation (CTR) by energized electrons escaping the target \cite{Liao:prl:116:205003}. In under- or near-critical plasmas,
it has been found experimentally that THz radiation can originate from CTR \cite{Leemans:prl:91:074802} or linear mode conversion of Langmuir waves excited in nonuniform density
profiles \cite{Liao:prl:114:255001}, both mechanisms leading to relatively low conversion efficiencies ($\eta \sim 10^{-6}$). 

In this Letter, we show by means of 3D particle-in-cell (PIC) simulations that gaseous targets driven at intensities $>10^{19}\,\mathrm{W/cm}^{2}$ by two-color pulses in the blowout wakefield
regime can also provide efficient ($\eta > 10^{-4}$) THz sources. Our study reveals that CTR can largely prevail over photoionization-induced radiation (PIR), yielding unprecedented $100\,\mathrm{GV/m}$ THz field strengths in gases. The simulation results are analyzed in light of the CTR theory and a simplified model of a radiating electron bunch
exiting into vacuum. Moreover, we derive an analytical formula for PIR that takes into account the nonlinear density modulations associated with the wakefield. Finally, we assess the dependence of the THz emission on the gas parameters.

\begin{figure}[ht!]
\includegraphics[width=\columnwidth]{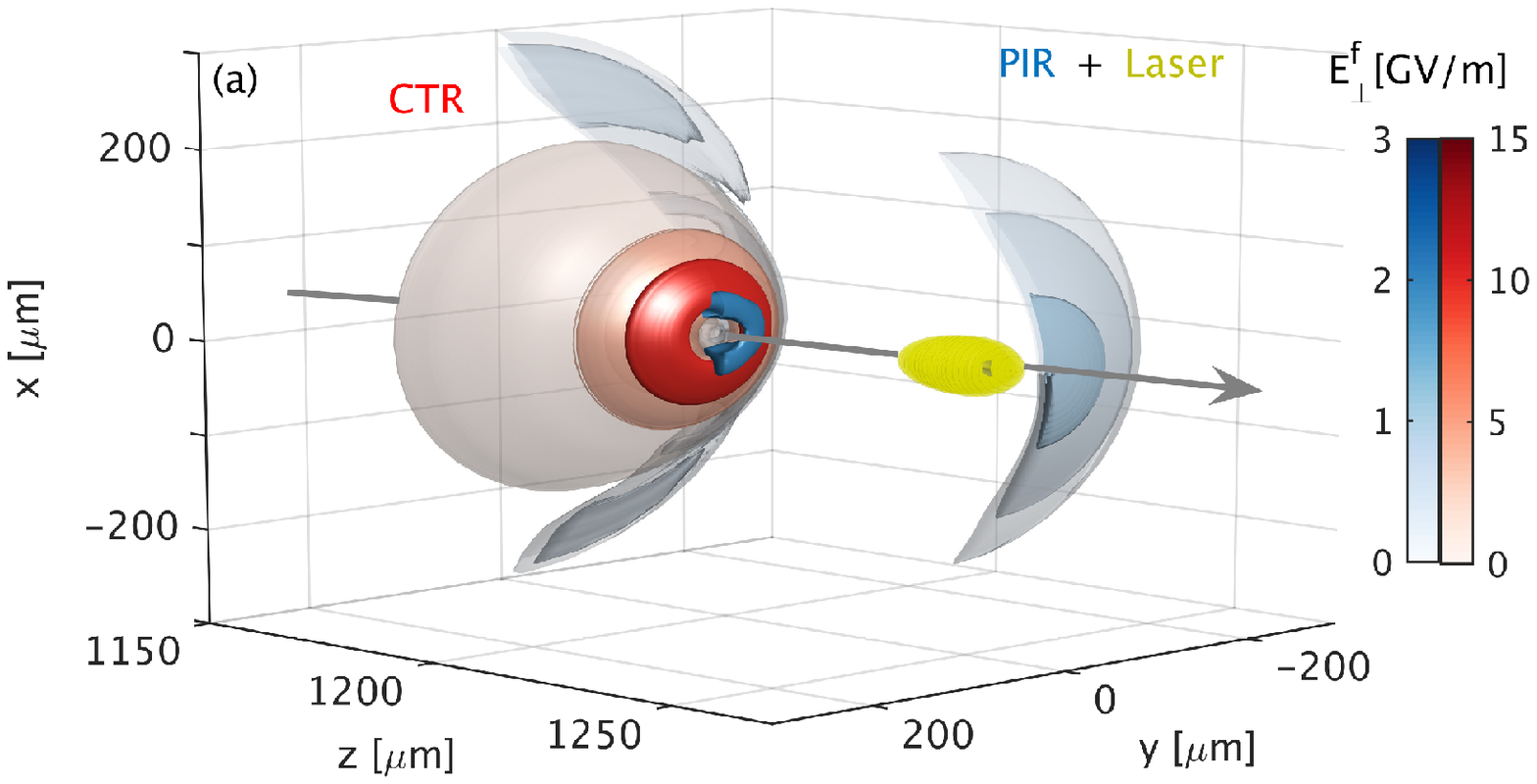}\\
\vspace{0.2cm}
\includegraphics[width=\columnwidth]{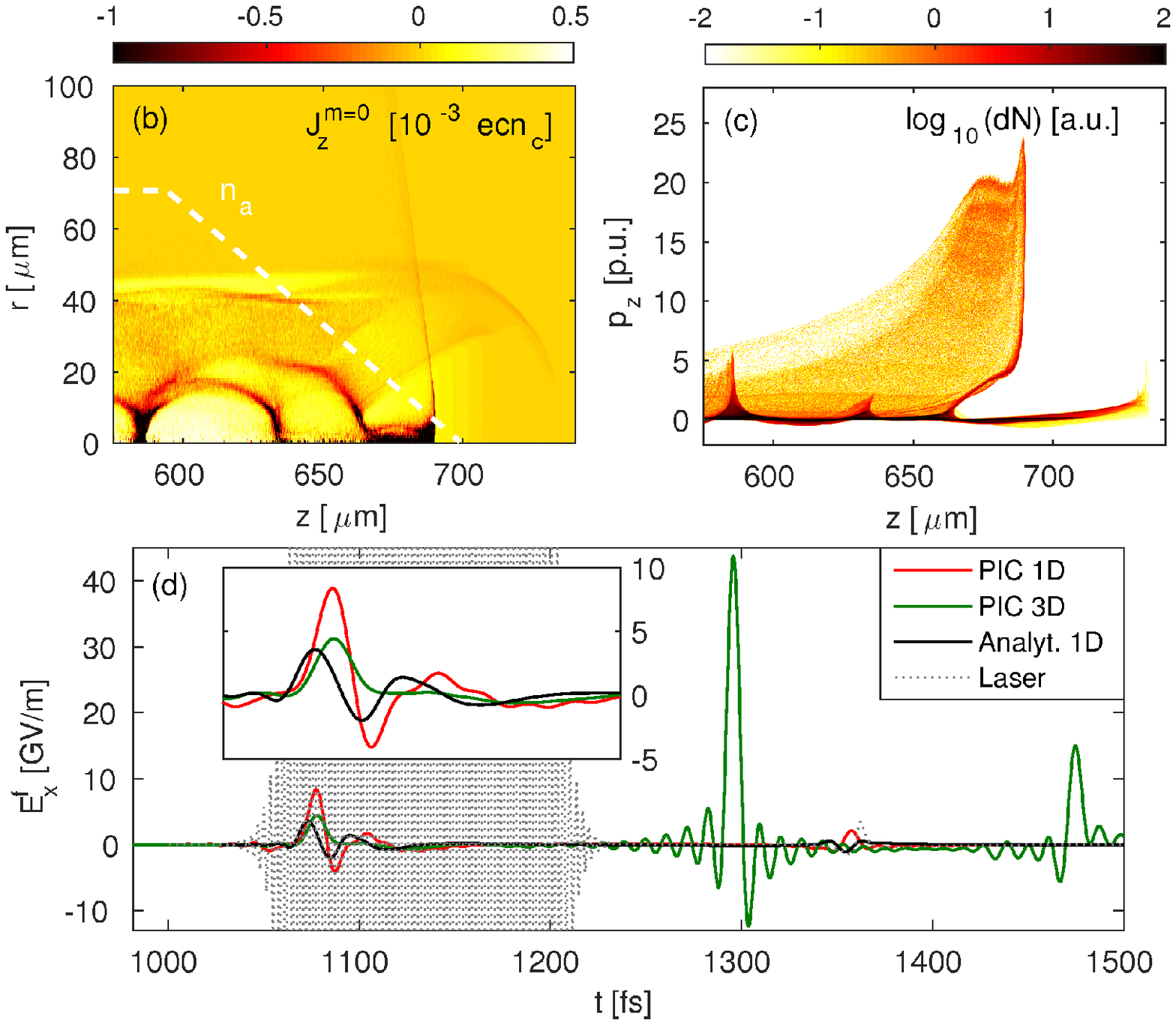}
\caption{THz field emission from a He gas target of $2.4\times 10^{17}\,\mathrm{cm}^{-3}$ atomic density driven by a $2.2\times 10^{19}\,\mathrm{W/cm}^{2}$, $35\,\mathrm{fs}$ two-color laser pulse. (a) 3D isosurfaces of the filtered ($\nu < 90\,\mathrm{THz}$) transverse field ($E_\perp^f$) in vacuum at $500\,\mu \mathrm{m}$ from the plasma boundary. Radially polarized ($m=0$) modes are fully displayed (red colormap); laser-polarized ($m=1$) ones are shown as half-caps for better readability (blue colormap). The yellow isosurface delineates the laser pulse envelope (propagating along the grey arrow) for a normalized field strength $a=2/3$ ($E_\perp = 2140\,\mathrm{GV/m}$). (b) Snapshot of the $m=0$ component of the longitudinal current ($J_z$) at time $t=2.55$ ps when the laser pulse exits the plasma-vacuum interface (atomic density is visualized by a white dashed line). (c) Electron number ($dN$ in $\log_{10}$ scale) in the ($z,p_z$) phase space at the same instant. (d) Time history of the on-axis $E_x^f$ field at $300\,\mu\mathrm{m}$ inside the plasma, as given by the 3D (green curve) and 1D (red curve) PIC simulations, and the solution of Eq.~\eqref{soluce} (black curve). The grey dashed curve represents the laser region.} 
\label{Fig1}
\end{figure}

Our 3D PIC simulations are performed using the \textsc{calder-circ} code \cite{Lifschitz:jcp:228:1803}. In the baseline simulation, a $5\,\mathrm{J}$ laser pulse, linearly polarized along
$x$ and propagating along the $z$ axis, is focused into a gas target of helium (He) with initial atomic density $n_a = 2.4\times 10^{17}\,\rm{cm}^{-3}$, $400\,\mu\mathrm{m}$ length ($L_p$),
and shaped with $100\,\mu\mathrm{m}$ long density ramps on both sides to mimic the conditions met in gas-jet experiments. In \textsc{calder-circ}, the electromagnetic fields are discretized on a $(r,z)$ grid and decomposed over a reduced set of Fourier angular modes ($\propto e^{i m\theta}$) around the $z$ axis. The fundamental $m=0$ mode corresponds to axisymmetric fields such as the radially polarized ones. The $m=1$ mode contains non-axisymmetric fields, including the $x$-polarized laser field. In order to optimize THz emissions \cite{Kim:np:2:605,Babushkin:njp:13:123029}, we consider a two-color laser field composed of a fundamental pulse with carrier wavelength $\lambda_0 \equiv 2 \pi c/\omega_0 = 1\,\mu\mathrm{m}$ ($\omega_0$ is the laser angular frequency and $c$ the velocity of light) and its second harmonic, shifted by a relative phase of $\pi/2$. The $2\omega_0/\omega_0$ intensity ratio is $10\%$ for a total laser intensity $I_0 = 2.2 \times 10^{19}\, \mathrm{W/cm}^2$, corresponding to a normalized field strength $a_0 \equiv e E_0/m_e \omega_0 c =4$ ($E_0$ is the electric field strength, $e$ the electron charge, and $m_e$ the electron mass). The laser harmonics have Gaussian profiles both in space and time with equal initial widths $w_0 = 20\,\rm{\mu m}$ and durations $\tau_0 = 35\,\mathrm{ fs}$ (FWHM). This setup fulfills the conditions for efficient electron blowout ($w_0 \omega_{pe}/c \approx 2 \sqrt{2 a_0 \ln 2}$, $\tau_0 \approx 2 w_0 / 3 c$ with $a_0 \gtrsim 4$, where $\omega_{pe}$ is the electron plasma frequency, see \cite{Lu:pop:13:056709}).

The THz fields are extracted by filtering the total field spectrum below a cut-off frequency $\omega_\mathrm{co} = 0.3\,\omega_0$ ($\nu_\mathrm{co} \equiv \omega_\mathrm{co}/2\pi = 90\,\mathrm{THz}$). Attention is paid to the transmitted THz fields only, as they usually prevail over the backscattered components in gases \cite{Debayle:pra:91:041801}. Inspection of the THz field in vacuum shows that the transverse field strength ($E_\perp$) exceeds the longitudinal one ($E_z$) by one order of magnitude. This invites us to restrict our analysis to $E_\perp$, whose PIR and CTR components can be discriminated through direct angular expansion: the PIR field is polarized along the laser field \cite{Gonzalez:sr:6:26743}, and so is described by the $m=1$ mode. By contrast, the wakefield-driven electron bunch is essentially axisymmetric, hence the resulting CTR (radially polarized) is mainly contained in the $m=0$ mode.

Figure~\ref{Fig1}(a) displays a set of isosurfaces of the PIR (blue colormap) and CTR (red colormap) electric fields at a distance of $500\,\mu\mathrm{m}$ from the rear side of the
plasma. The rightward-propagating laser pulse is visualized by the yellow isosurface.
A primary PIR burst occurs at a distance of $\sim 20\,\mu\mathrm{m}$ in front of the laser peak, reaching a maximum amplitude of $\sim 1\,\mathrm{GV/m}$ on axis and carrying a total energy of $1.3\,\mu\mathrm{J}$.  About one plasma wavelength behind the laser pulse, a radially polarized burst produces the maximum THz field $\sim 15\,\mathrm{GV/m}$, corresponding to a $\sim 160\,\mu\mathrm{J}$ energy. The location and the hollow conical shape of this emission are consistent with CTR by electrons
accelerated in the laser wakefield, as justified below. In the present UHI conditions, this strongly nonlinear plasma wave takes the form of a succession of ion cavities due to radial expulsion of the plasma electrons \cite{Lu:pop:13:056709}. The intense burst evidenced in Fig.~\ref{Fig1}(a) is emitted when the electron bunch that has been trapped into the first cavity exits into vacuum. This scenario is supported by Figs.~\ref{Fig1}(b,c), which show (b) the $m=0$ component of the longitudinal electron current density ($J_z$) and (c) the electron $(z,p_z)$ phase space at time $t=2.55\,\mathrm{ps}$. The strong peak in $J_z$ seen at the foot of the density down-ramp ($z\simeq 690\,\mu\mathrm{m}$) corresponds to a high-energy ($p_z \gtrsim 25\,m_ec$) electron bunch about to exit the plasma. Subsequent cavities also accelerate a few electron packets, yet at lower energies and densities. Finally, Fig.~\ref{Fig1}(a) reveals a secondary non-axisymmetric signal on top of the CTR. This emission, less collimated but more intense ($\sim 3\,\mathrm{GV/m}$, corresponding to an energy of $\sim 2.6\,\mu\mathrm{J}$) than that occurring in the laser front pulse, stems from the coupling of the transverse photocurrents and the strong density oscillations accompanying the wakefield.

To gain insight into the PIR, we plot in Fig.~\ref{Fig1}(d) the time history of the on-axis filtered $E_x^f$ field at a depth of $300\,\mu\mathrm{m}$ inside the plasma (green curve), and we
compare it with the result of a 1D PIC simulation using the same parameters (red curve). Relatively good agreement (within a factor $\sim 2$) is found between the
1D and 3D curves during the laser pulse [see also inset of Fig.~\ref{Fig1}(d)], both showing peak fields of $\sim 4-8\,\mathrm{GV/m}$. Outside the plasma, diffraction causes this primary THz emission to weaken as it propagates in vacuum [Fig.~\ref{Fig1}(a)]. As detailed in the Supplemental Material \cite{SupMat}, an expression for the 
PIR field can be derived in a 1D geometry by assuming an unperturbed laser pulse moving at $c$ and a stationary plasma wave in the co-moving coordinate system
($\xi=z-ct$ and $s=t$). By introducing the vector potential $A_x=A_L(\xi)+\delta A_x(\xi,s)$ and transverse momentum $p_x=p_L(\xi)+e\delta A_x(\xi,s)$ ($|\delta A_x|\ll |A_L|$),
the 1D wave equation of the radiated potential vector reads
\begin{equation}
\label{eqtransf}
-\partial_{ss} \delta A_x+2c\partial_{s\xi}\delta A_x=\frac{en_e}{m_e\varepsilon_0\gamma_e}(p_L+e\delta A_x),
\end{equation} 
where $\epsilon_0$ is the permittivity of vacuum, $n_e$ is the electron density and $\gamma_e$ is the Lorentz factor associated with the electron velocity $v_e$. For a laser pulse located
in the half-plane $z<ct$ ($\xi<0$) and entering the plasma at time $t=0$, Eq.~\eqref{eqtransf} has the following solution:
\begin{align}
 \delta A_x=&\frac12\sqrt{\frac{1}{m_e\varepsilon_0c^2}}\int_0^\xi d\xi^\prime\frac{n_ep_L}{\gamma_e}(\xi')\sqrt{\frac{2cs+\xi-\xi^\prime}{\int_\xi^{\xi^\prime}\frac{n_e}{\gamma_e} d\xi^{\prime\prime}}} \nonumber\\
&\times J_1\left[\sqrt{\int_\xi^{\xi^\prime}\frac{e^2n_e}{m_e\varepsilon_0 \gamma_e} d\xi^{\prime\prime}(2cs+\xi-\xi^\prime)}\right] \,, \label{soluce}
\end{align}
where $J_1(x)$ is the Bessel function of the first kind. The electron density is computed numerically from the standard wakefield equation, supplemented with a photoionization
source \cite{SupMat}. Equation \eqref{soluce} describes the coupling between the transverse photocurrents, which mediate the usual PIR during the laser pulse \cite{Debayle:oe:22:13691}, 
and the density modulations associated with both photoionization and the nonlinear laser wakefield. This formula correctly reproduces the 1D PIC result during the laser pulse [see inset
of Fig.~\ref{Fig1}(d)]. Also, due to the interplay of PIR and wakefield, THz bursts occur at each density peak with a $\sim 200\,\mathrm{fs}$ period, as seen at time $t \simeq 1360\,\mathrm{fs}$ in both the 1D PIC and theoretical curves. In the 3D simulation, the corresponding emission occurs a bit earlier ($t \simeq 1300\,\mathrm{fs}$) and at a much higher amplitude ($\sim 40\,\mathrm{GV/m}$
vs $\sim 2\,\mathrm{GV/m}$ in 1D); this is explained by differences in the dynamics and shape of the 1D and 3D plasma waves, the latter being subject to complete electron blowout. This signal vanishes in a 3D preionized plasma, hence demonstrating the role of photocurrents. It still prevails over the primary PIR burst in vacuum at $500\,\mu\mathrm{m}$ from the plasma, although being more strongly reduced by diffraction down to $3$ GV/m [Fig.~\ref{Fig1}(a)].

\begin{figure}
\includegraphics[width=\columnwidth]{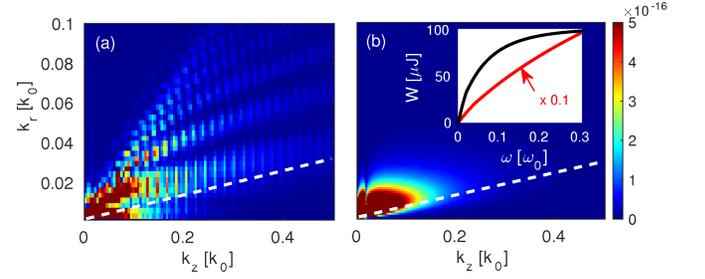}
\caption{2D $(k_z,k_r)$ spectrum of the axisymmetric field $E_\perp^f$ in vacuum at $500\,\mu\mathrm{m}$ from the plasma: (a) 3D PIC simulation results; (b) CTR theory [Eq.~\eqref{CTR}] for $\gamma_e=15$ and $N_e=8.9 \times 10^8$. 
The white dashed lines indicate the direction of maximum emission, $\theta_\mathrm{max} = \gamma_e^{-1}$. The inset in (b) shows the emitted field energy [$\mu\mathrm{J}$] versus frequency according to Eq.~\eqref{CTR} for $n_a = 2.4\times 10^{17} \,\mathrm{cm}^{-3}$ (black curve) and $n_a = 9.7\times 10^{17} \, \mathrm{cm}^{-3}$ (red curve).
In the latter case discussed in Fig.~\ref{Fig4}(d), the electron bunch parameters are $\gamma_e =125$ and $N_e = 1.2 \times 10^9$.}
\label{Fig2}
\end{figure}

We now turn to the analysis of the brightest, radially polarized signal measured in vacuum [Fig.~\ref{Fig1}(a)]. To prove that it mainly arises from CTR by wakefield-driven electrons, we confront its 2D energy spectrum to that predicted from a point-like monoenergetic electron bunch exiting perpendicularly to the plasma surface \cite{Garibian:jetp:6:1079,Jackson:CE:75,Zheng:pop:10:2994,Schroeder:pre:69:016501}:
\begin{align}
\frac{d^2 W_e}{d \omega d \Omega}  &=  \frac{c N_e^2e^2 \sin^2 \theta \cos^2 \theta}{\pi^2 v_e^2} \frac{\beta_e^4}{(1-\beta_e^2 \cos^2 \theta)^2}  \nonumber \\ & \times \left | \frac{(\epsilon -1)(1 - \beta_e^2 - \beta_e \sqrt{\epsilon - \sin^2 \theta})}{(\epsilon \cos \theta + \sqrt{\epsilon - \sin^2 \theta})(1 - \beta_e \sqrt{\epsilon - \sin^2 \theta})} \right |^2 \,. \label{CTR}
\end{align}
Here, $d^2W_e/d \omega d \Omega$ is the radiated energy density per units of angular frequency ($\omega$) and solid angle ($\Omega$), $\theta$ is the angle between the propagation
axis and the observer, $N_e$ is the number of electrons inside the bunch, $v_e = \beta_e c$ is their velocity, and $\epsilon = 1-\omega_{pe}^2/\omega^2$ is the plasma dielectric
function. The assumption of a point-like electron bunch holds provided that the bunch size is much smaller than the radiation wavelength, in which case the emission is coherent \cite{Schroeder:pre:69:016501}.  
Equation (\ref{CTR}) can be recast in terms of the longitudinal ($k_z$) and transverse ($k_r$) wave numbers using $\theta = \arctan(k_r / k_z)$ and $\omega = c \sqrt{k_z^2 + k_r^2}$. Figures \ref{Fig2}(a,b) show the THz
spectra computed from (a) the 3D PIC simulation and (b) from Eq.~\eqref{CTR} using the mean values $\gamma_e = 1/\sqrt{1-\beta_e^2} = 15$ and $N_e = 8.9 \times 10^8$ that best fit
the electron bunch issued from the first wakefield bucket [with $p_z \geq 5\,m_ec$ in Fig. \ref{Fig1}(c)]. Despite the crude simplifications of Eq.~\eqref{CTR}, the two spectra fairly agree in intensity and shape: both present
a maximum emission along $\theta_\mathrm{max} \simeq \gamma_e^{-1}$ (white dashed line) with a cutoff frequency $\omega_\mathrm{max} \simeq \gamma_e \omega_{pe} \approx 0.3\,\omega_0$,
as expected from CTR by relativistic electrons \cite{Jackson:CE:75}. The PIC spectrum, however, differs from the theoretical one by additional weaker emissions at larger angles and spectral modulations
separated by $\Delta k_r  \simeq \omega_{pe}/c$, which are ascribed to radiation by lower-energy electron bunches produced in the second and third wakefield buckets.
The inset of Fig.~\ref{Fig2}(b) plots the theoretical radiated energy [$\mu\mathrm{J}$] as a function of the frequency (black curve). In the THz frequency
range $\omega < 0.3\, \omega_0$, we obtain a total energy of $\sim 100\,\mu\mathrm{J}$, comparable with the $\sim 160\,\mu\mathrm{J}$ yield measured in the simulation. 

\begin{figure}
\includegraphics[width=\columnwidth]{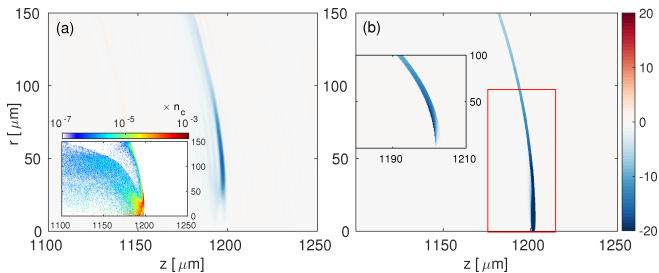}
\caption{(a) 2D $(z,r)$ map of the azimuthal magnetic field, $cB_\theta$ [$\mathrm{GV/m}$], extracted at time $t=4\,\mathrm{ps}$ from the 3D PIC simulation of Fig.~\ref{Fig1}. The electron bunch and the plasma-vacuum interface are located at $z\approx 1200\,\mu\mathrm{m}$ and $z = 700\,\mu\mathrm{m}$, respectively.
The inset  shows the electron density $n_e$ [$n_c$].
(b) Same quantity as in (a) but given by the Biot-Savart law [Eq.~\eqref{Biot-Savart}].
The electron bunch parameters are detailed in the text.
The inset displays the field emitted at the plasma-vacuum surface, obtained by subtracting the asymptotic proper field of the electron bunch to the total Biot-Savart
field. All fields in (a) and (b) are filtered in the frequency range $\nu < 90\,\mathrm{THz}$.
}
\label{Fig3}
\end{figure}

In addition to CTR, the radially polarized THz spectrum measured outside the plasma in the PIC simulation also includes the proper field of the electron bunch, which is not described by Eq.~\eqref{CTR}. This field is of electrostatic character in the bunch rest frame, and should thus be discarded when evaluating the source efficiency in emitting purely electromagnetic THz radiations. To model the space-time field distribution resulting from both the plasma-boundary crossing and subsequent propagation of the electron bunch, we make use of the generalized Bio-Savart law \cite{Bellotti:ajp:64:569}:
\begin{equation}
\vec{B}(\vec{r},t) = \int d^3 r' \, \left\{ \frac{[ \vec{J} (\vec{r'},t') ]}{cR^2} + \frac{1}{c^2 R} \left[ \frac{\partial \vec{J} (\vec{r'},t')}{\partial t' } \right] \right \} \times \frac{\vec{R}}{R} \,,
\label{Biot-Savart}
\end{equation}
where $\vec{B}$ is the magnetic field, $\vec{J}$ is the current density, $\vec{r}$ is the observer's position, $\vec{R} \equiv \vec{r} - \vec{r}'$ and the square brackets denote an
evaluation at the retarded time $t' \equiv t - R/c$. Equation \eqref{Biot-Savart} is computed for a monoenergetic electron bunch of zero radius and finite length $L_e$, moving at
constant velocity along the $z$-axis \cite{SupMat}. Transition radiation arises from assuming that the beam emerges into vacuum through the plasma surface, which implies that
the plasma behaves as a perfect conductor. The specific features of the CTR-like and proper fields are discussed in \cite{SupMat}.
Figures \ref{Fig3}(a,b) confront (b) the result of Eq.~\eqref{Biot-Savart} to (a) the axisymmetric $B_\theta$ field measured in the 3D simulation when the main electron bunch
has propagated $500\,\mu\mathrm{m}$ beyond the interface. In Fig.~\ref{Fig3}(b), we use the parameter values $\gamma_e=15$, $N_e=8.9 \times 10^8$ and $L_e=1.5\,\mu\mathrm{m}$. Good agreement is found outside the bunch ($r \gtrsim 30\,\mu\mathrm{m}$) between the two maps of $B_\theta$ filtered in the THz band $\nu < 90\,\mathrm{THz}$, both in amplitude and spatial shape. The main discrepancy is found inside the bunch, for which Eq.~\eqref{Biot-Savart} overestimates the simulated field due to the assumed zero radius of the bunch, whereas the latter diverges to some extent in the simulation [inset of Fig.~\ref{Fig3}(a)]. To isolate the CTR in our calculation, we subtract the asymptotic proper field of the bunch \cite{SupMat} to the total field, and plot the result in the inset of Fig.~\ref{Fig3}(b). From comparison of this graph with the total field distributions of Figs.~\ref{Fig3}(a,b), it appears that most of the off-axis (axisymmetric) THz emission ($r \gtrsim 30 \,\mu\mathrm{m}$) indeed originates from the plasma-vacuum interface.

\begin{figure}[ht]
\includegraphics[width=\columnwidth]{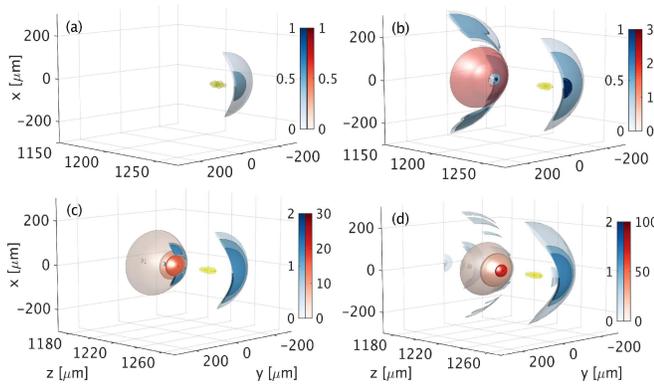} 
\caption{3D isosurfaces of the THz transverse field ($E_\perp^f$) at $500\,\mu \mathrm{m}$ from the plasma-vacuum interface for different gas parameters: 
(a) $n_a = 6\times 10^{16}\,\mathrm{cm}^{-3}$; (b) $n_a = 2.4\times 10^{17}\,\mathrm{cm}^{-3}$ with a steplike rear plasma boundary; (c,d) $n_a=9.7 \times 10^{17}\,\mathrm{cm}^{-3}$ with (c) a steplike rear plasma boundary and (d) a $100\,\mu\mathrm{m}$ long density ramp. Red (blue) colormaps correspond to axisymmetric (resp. non-axisymmetric) fields.}
\label{Fig4}
\end{figure}

The CTR yield evidently depends on the efficiency of the wakefield acceleration, and is therefore sensitive to the gas parameters. As shown in Fig.~\ref{Fig4}(a), when decreasing
the gas density to $n_a = 6\times 10^{16}\,\mathrm{cm}^{-3}$, the energy and number of the escaping electrons significantly drop ($\gamma_e \simeq 4$, $N_e \simeq 10^7$), which in turn reduces the CTR ($\sim 0.1 \,\mathrm{GV/m}$) much below the PIR level ($\sim 1\,\mathrm{GV/m}$). Returning to the reference configuration but changing the density ramp at the rear side of the gas to a sharp gradient [Fig.~\ref{Fig4}(b)], the PIR signal is essentially unmodified [compare with Fig.~\ref{Fig1}(a)], while the CTR signal is significantly weakened ($\sim 2\,\mathrm{GV/m}$) due to an order of magnitude reduced $N_e$. This pinpoints the beneficial role of the $100$-$\mu\mathrm{m}$ density down-ramp in our reference setup, which promotes gradient injection \cite{Bulanov:pre:58.R5257}. Similarly, Figs.~\ref{Fig4}(c,d) illustrate the case of a $4$ times denser gas ($n_a = 9.7\times 10^{17}\,\mathrm{cm}^{-3}$) with (c) a sharp rear boundary and (d) a $100\,\mu\mathrm{m}$-long ramp. Both setups lead to stronger wakefields, still in the blowout regime. As the plasma length remains much shorter than the dephasing length ($L_p = 400\,\mu\mathrm{m} \ll L_d = 2 \omega_0^2 w_0/3 \omega_{pe}^2 \simeq 3.3\,\mathrm{cm}$), there result electron bunches of larger charge and energy, up to $N_e \simeq 1.2\times10^9$ and $\gamma_e \simeq 125$, thus generating stronger CTR than at $n_a = 2.7 \times 10^{17}\,\mathrm{cm}^{-3}$. Again, the presence of a density ramp proves advantageous, augmenting the CTR field (energy) from $\sim 30\,\mathrm{GV/m}$ ($\sim 750\,\mu\mathrm{J}$) to $\sim 100\,\mathrm{GV/m}$ ($\sim 1.36\,\mathrm{mJ}$). The main electron bunch supplies a total THz energy consistent with the theoretical CTR spectrum ($\sim 0.92\,\mathrm{mJ}$) displayed with a red curve in the inset of Fig.~\ref{Fig2}(b). Note that the $\sim$ 10-fold increase in the CTR energy yield and field strength is consistent with the linear scaling in $\gamma_e$ expected from CTR theory \cite{Durand:prd:11:89,Jackson:CE:75}.
The relatively weak enhancement ($\times 2$) of the primary PIR field illustrates its nontrivial dependency on the interferences caused by ionization events and the local relative phase between the two colors \cite{Babushkin:njp:13:123029}.

In summary, by means of full-scale 3D PIC simulations, we have evidenced the sequential production of intense THz bursts using two-color UHI ultrashort laser pulses interacting with He gases of sub-millimeter lengths and $>10^{17}\,\mathrm{cm}^{-3}$ atomic densities. Following a primary THz burst induced by photocurrents, CTR at the rear plasma boundary by wakefield-driven relativistic electrons can generate THz pulses of $\sim 100\,\mathrm{GV/m}$ field strengths and $> 1\,\mathrm{mJ}$ energies using relatively modest laser parameters ($5\,\mathrm{J}$ in energy, $2.2\times 10^{19}\,\mathrm{W/cm}^{2}$ in intensity). We have obtained an analytical formula that captures the on-axis patterns of the $\mathrm{GV/m}$-level PIR predicted by 1D and 3D simulations. Furthermore, analytical CTR models satisfactorily match the simulated radiation in terms of spectral and field distributions. Finally, we have gauged the sensitivity of the CTR and PIR to the interaction setup by varying the gas density and density scale length. Further studies should focus on the electron acceleration stage to provide even more powerful THz sources.
 
The authors acknowledge GENCI, France for awarding us access to the supercomputer CURIE using Grant $\#$ 2016-057594.

\bibliography{references_prl}

\end{document}